\documentclass[journal]{IEEEtran}

\usepackage{cite}
\usepackage{amsmath,amssymb,amsfonts,amsthm}
\usepackage{algorithmic}
\usepackage{graphicx}
\usepackage{textcomp}
\usepackage{xcolor}
\usepackage{cases}
\usepackage[utf8]{inputenc}
\usepackage[english]{babel}

\usepackage[normalem]{ulem}

\usepackage{epstopdf}

\usepackage{enumitem}

\usepackage{hyperref}
\hypersetup{colorlinks=true,linkcolor=black,urlcolor=blue,}

\usepackage{dirtytalk}

\usepackage[caption=false]{subfig} 
\makeatletter
\renewcommand{\fnum@figure}{Fig. \thefigure}
\makeatother

\usepackage[acronym,toc,shortcuts]{glossaries}
\newacronym{phd}{Ph.D.}{Doctor of Philosophy}
\newacronym{SA}{SA}{Simulated Annealing}
\newacronym{VLC}{VLC}{visible light communications}
\newacronym{RF}{RF}{radio frequency}
\newacronym{V2V}{V2V}{vehicle-to-vehicle}
\newacronym{V2X}{V2X}{vehicle-to-everything}
\newacronym{V2I}{V2I}{vehicle-to-infrastructure}
\newacronym{B5G}{B5G}{beyond-fifth generation}
\newacronym{LED}{LED}{light emitting diode}
\newacronym{OMA}{OMA}{orthogonal multiple access}
\newacronym{FDMA}{FDMA}{frequency-division multiple-access}
\newacronym{TDMA}{TDMA}{time-division multiple-access}
\newacronym{CDMA}{CDMA}{code-division multiple-access}
\newacronym{OFDMA}{OFDMA}{orthogonal frequency-division multiple-access}
\newacronym{OFDM}{OFDM}{orthogonal frequency-division multiplexing}
\newacronym{WDMA}{WDMA}{wavelength-division multiple-access}
\newacronym{NOMA}{NOMA}{non-orthogonal multiple access}
\newacronym{PD-NOMA}{PD-NOMA}{power-domain NOMA}
\newacronym{CD-NOMA}{CD-NOMA}{code-domain NOMA}
\newacronym{SC}{SC}{superposition coding}
\newacronym{SIC}{SIC}{successive interference cancellation}
\newacronym{BS}{BS}{base station}
\newacronym{QoS}{QoS}{quality-of-service}
\newacronym{NP}{NP}{non-deterministic polynomial-time}
\newacronym{DCO-OFDM}{DCO-OFDM}{direct-current biased optical-OFDM}
\newacronym{DCO-OFDMA}{DCO-OFDMA}{direct-current biased optical-OFDMA}
\newacronym{DC}{DC}{direct current}
\newacronym{ITU}{ITU}{international telecommunication union}
\newacronym{FoV}{FoV}{field-of-view}
\newacronym{CSI}{CSI}{channel state information}
\newacronym{LACO-OFDM}{LACO-OFDM}{layered asymmetrically clipped optical OFDM}
\newacronym{ACO-OFDM}{ACO-OFDM}{asymmetrically clipped optical OFDM}
\newacronym{FR}{FR}{frequency reuse}
\newacronym{EAs}{EAs}{evolutionary algorithms}
\newacronym{C-LiAN}{C-LiAN}{centralized light access network}
\newacronym{AP}{AP}{access point}
\newacronym{PD}{PD}{photo-diode}
\newacronym{SINR}{SINR}{signal-to-noise-interference ratio}
\newacronym{LoS}{LoS}{line-of-sight}
\newacronym{AWGN}{AWGN}{additive white Gaussian noise}
\newacronym{SNR}{SNR}{signal-to-noise ratio}
\newacronym{NLUPA}{NLUPA}{next-largest-difference user-pairing algorithm}
\newacronym{D-NLUPA}{D-NLUPA}{divide-and-next-largest-difference user-pairing algorithm}
\newacronym{NAICS}{NAICS}{network-assisted interference cancellation and suppression}
\newacronym{LTE}{LTE}{long term evolution}
\newacronym{3GPP}{3GPP}{3rd Generation Partnership Project}
\newacronym{CR}{CR}{cognitive radio}
\newacronym{2D}{2D}{two-dimension}
\newacronym{GP}{GP}{gradient projection}
\newacronym{umMTC}{umMTC}{ultra-massive machine-type communication}
\newacronym{mMTC}{mMTC}{massive machine-type communication}
\newacronym{IoE}{IoE}{internet-of-everything}
\newacronym{IoUT}{IoUT}{internet-of-underwater-things}
\newacronym{ZF}{ZF}{zero-forcing}
\newacronym{NLIP}{NLIP}{non-linear integer programming}
\newacronym{DP}{DP}{dynamic programming}
\newacronym{MIMO}{MIMO}{multiple-input multiple-output}
\newacronym{TS}{TS}{Tabu-search}
\newacronym{THz}{THz}{Terahertz}
\newacronym{MISO}{MISO}{multiple-input single-output}
\newacronym{SIMO}{SIMO}{single-input multiple-output}
\newacronym{EE}{EE}{energy efficiency}
\newacronym{VR}{VR}{virtual reality}
\newacronym{XR}{XR}{extended reality}
\newacronym{5G}{5G}{fifth generation}
\newacronym{6G}{6G}{sixth generation}
\newacronym{NR}{NR}{new radio}
\newacronym{mmWave}{mmWave}{millimeter-wave}
\newacronym{FD}{FD}{full-duplex}
\newacronym{CNOMA}{CNOMA}{cooperative NOMA}
\newacronym{ABF}{ABF}{analog beamforming}
\newacronym{BF}{BF}{beamforming}
\newacronym{DF}{DF}{decode-and-forward}
\newacronym{AF}{AF}{amplify-and-forward}
\newacronym{CF}{CF}{compress-and-forward}
\newacronym{SPS}{SPS}{single-phase shifter}
\newacronym{PS}{PS}{phase shifter}
\newacronym{PA}{PA}{power amplifier}
\newacronym{NLoS}{NLoS}{non-line-of-sight}
\newacronym{ULA}{ULA}{uniform linear array}
\newacronym{SI}{SI}{self-interference}
\newacronym{MA}{MA}{multiple access}
\newacronym{1G}{1G}{first generation}
\newacronym{2G}{2G}{second generation}
\newacronym{3G}{3G}{third generation}
\newacronym{4G}{4G}{fourth generation}
\newacronym{M2M}{M2M}{machine-to-machine}
\newacronym{IoT}{IoT}{internet-of-things}
\newacronym{IMT}{IMT}{International Mobile Telecommunications}
\newacronym{SE}{SE}{spectral efficiency}
\newacronym{MMF}{MMF}{Maximin Fairness}
\newacronym{SR}{SR}{sum rate}
\newacronym{WSR}{WSR}{weighted sum rate}
\newacronym{D-NOMA}{D-NOMA}{dynamic-NOMA}
\newacronym{GSM}{GSM}{global system for mobile telecommunications}
\newacronym{IS-95}{IS-95}{Interim Standard 95}
\newacronym{SMS}{SMS}{short-message service}
\newacronym{CoMP}{CoMP}{coordinated multi-point}
\newacronym{MU-MIMO}{MU-MIMO}{multi-user MIMO}
\newacronym{MAC}{MAC}{multiple-access channel}
\newacronym{BC}{BC}{vector-broadcast channel}
\newacronym{CSIT}{CSIT}{channel state information at the transmitter}
\newacronym{DPC}{DPC}{dirty paper coding}
\newacronym{ZF-DPC}{ZF-DPC}{zero-forcing DPC}
\newacronym{BD}{BD}{block-diagonalization}
\newacronym{ICI}{ICI}{inter-cell interference}
\newacronym{MMSE}{MMSE}{minimum mean square error}
\newacronym{LTE-A}{LTE-A}{long term evolution-advanced}
\newacronym{MUST}{MUST}{multi-user superposition transmission}
\newacronym{SISO}{SISO}{single-input single-output}
\newacronym{LDS-CDMA}{LDS-CDMA}{low-density spreading CDMA}
\newacronym{LDS-OFDM}{LDS-OFDM}{low-density spreading OFDM}
\newacronym{SCMA}{SCMA}{sparse code multiple access}
\newacronym{MUSA}{MUSA}{multi-user sharing access}
\newacronym{SAMA}{SAMA}{successive interference cancellation amenable multiple access}
\newacronym{PDMA}{PDMA}{pattern division multiple access} 
\newacronym{BOMA}{BOMA}{building block sparse-constellation based orthogonal multiple access}  
\newacronym{LPMA}{LPMA}{lattice partition multiple access} \newacronym{OOK}{OOK}{on-off keying} 
\newacronym{M-PAM}{M-PAM}{M-ary pulse-amplitude modulation} 
\newacronym{M-PPM}{M-PPM}{M-ary pulse-position modulation} \newacronym{MSM}{MSM}{multiple-subcarrier modulation}
\newacronym{IM/DD}{IM/DD}{intensity modulation and direct detection}
\newacronym{RC}{RC}{repetition code}
\newacronym{SM}{SM}{spatial multiplexing}
\newacronym{SMOD}{SMOD}{spatial modulation}
\newacronym{BER}{BER}{bit error rate}
\newacronym{SER}{SER}{symbol error rate}
\newacronym{MFTP}{MFTP}{maximum flickering time period}
\newacronym{MU-MISO}{MU-MISO}{multi-user multi-input single-output}
\newacronym{MSE}{MSE}{minimum square error}
\newacronym{SPCA}{SPCA}{sequential parametric convex approximation}
\newacronym{WSMSE}{WSMSE}{weighted sum minimum square error}
\newacronym{OCDMA}{OCDMA}{optical code division multiple access}
\newacronym{SDMA}{SDMA}{space-division multiple access}
\newacronym{DMT}{DMT}{discrete multi-tone}
\newacronym{VLNs}{VLNs}{visible light networks}
\newacronym{VHO}{VHO}{Vertical handover}
\newacronym{RSS}{RSS}{received signal strength}
\newacronym{RSI}{RSI}{received signal intensity}
\newacronym{IA}{IA}{interference alignment}
\newacronym{BIA}{BIA}{blind interference alignment}
\newacronym{BBU}{BBU}{base-band unit}
\newacronym{SU}{SU}{secondary user}
\newacronym{PU}{PU}{primary user}
\newacronym{mMIMO}{mMIMO}{massive-MIMO}
\newacronym{UAV}{UAV}{unmanned aerial vehicle}
\newacronym{PHY}{PHY}{physical}
\newacronym{CF-mMIMO}{CF-mMIMO}{cell-free mMIMO}
\newacronym{LIS}{LIS}{large intelligent surfaces}
\newacronym{3-D MIMO}{3-D MIMO}{3-Dimensional MIMO}
\newacronym{RIS}{RIS}{reflecting intelligent surface}
\newacronym{BackCom}{BackCom}{backscatter communications}
\newacronym{UL}{UL}{uplink}
\newacronym{UE}{UE}{user equipment}
\newacronym{D2D}{D2D}{device-to-device}
\newacronym{FCC}{FCC}{Federal Communications Commission}
\newacronym{HAP}{HAP}{high altitude platform}
\newacronym{LAP}{LAP}{low altitude platform}
\newacronym{MEC}{MEC}{mobile edge computing}
\newacronym{NATO}{NATO}{North Atlantic Treaty Organization}
\newacronym{ML}{ML}{machine learning}
\newacronym{QML}{QML}{quantum machine learning}
\newacronym{DL}{DL}{deep learning}
\newacronym{DRL}{DRL}{deep Reinforcement learning}
\newacronym{RL}{RL}{Reinforcement learning}
\newacronym{MMA}{MMA}{minorization maximization algorithm}
\newacronym{MM}{MM}{majorization-minimization}
\newacronym{KKT}{KKT}{Karush–Kuhn–Tucker}
\newacronym{FDD}{FDD}{frequency division duplex}
\newacronym{SCA}{SCA}{sine-cosine algorithm}
\newacronym{AoD}{AoD}{angle-of-departure}
\newacronym{SDP}{SDP}{semi-definite programming}
\newacronym{SDR}{SDR}{semi-definite relaxation}
\newacronym{JT}{JT}{joint transmission}
\newacronym{CB}{CB}{coordinated beamforming}
\newacronym{RAMA}{RAMA}{relay-aided multiple access}
\newacronym{MRC}{MRC}{maximum ratio combining}
\newacronym{SWIPT}{SWIPT}{simultaneous wireless information and power transfer}
\newacronym{HetNets}{HetNets}{heterogeneous networks}
\newacronym{D.C.}{D.C.}{difference of convex}
\newacronym{GRPA}{GRPA}{gain ratio power allocation}
\newacronym{FPA}{FPA}{fixed power allocation}
\newacronym{CS}{CS}{Cuckoo Search}
\newacronym{HHO}{HHO}{Harris Hawks Optimizer}
\newacronym{PLC}{PLC}{power line communications}
\newacronym{HTT}{HTT}{harvest-then-transmit}
\newacronym{H-CRAN}{H-CRAN}{heterogeneous cloud radio access network}
\newacronym{RRHs}{RRHs}{remote radio heads}
\newacronym{IIoT}{IIoT}{Industrial IoT}
\newacronym{PAPR}{PAPR}{peak-to-average-power-ratio}
\newacronym{ANC}{ANC}{analog network coding}
\newacronym{FFR}{FFR}{fractional frequency reuse}
\newacronym{RGB}{RGB}{red-green-blue}
\newacronym{MAR}{MAR}{mobile augmented reality}
\newacronym{HD}{HD}{half-duplex}
\newacronym{CPU}{CPU}{central process unit}
\newacronym{RWP}{RWP}{Random Way-Point}
\newacronym{LC}{LC}{liquid crystal}
\newacronym{ADR}{ADR}{angle diversity receiver}
\newacronym{OWC}{OWC}{optical wireless communications}

\usepackage[linesnumbered,ruled,vlined]{algorithm2e}
\SetKw{KwBy}{by}

\usepackage{qtree}

\begin{document}

\title{Optimized Design of Joint Mirror Array and Liquid Crystal-Based RIS-Aided VLC systems}

\author{Omar~Maraqa and Telex M. N. Ngatched, \IEEEmembership{Senior Member,~IEEE}%

\thanks{The authors are with the Department of Electrical and Computer Engineering, McMaster University, Hamilton, Canada (e-mail: dr.omar.maraqa@gmail.com; ngatchet@mcmaster.ca). This work was supported in part by the Natural Sciences and Engineering Research Council of Canada (NSERC) through its Discovery program, and in part by McMaster University. (Corresponding author: Omar Maraqa). }%
\thanks{2023 IEEE.  Personal use of this material is permitted.  Permission from IEEE must be obtained for all other uses, in any current or future media, including reprinting/republishing this material for advertising or promotional purposes, creating new collective works, for resale or redistribution to servers or lists, or reuse of any copyrighted component of this work in other works. Digital Object Identifier: \href{https://ieeexplore.ieee.org/abstract/document/10183987}{10.1109/JPHOT.2023.3295350}}%
}


\maketitle

\begin{abstract}
Most studies of reflecting intelligent surfaces (RISs)-assisted visible light communication (VLC) systems have focused on the integration of RISs in the channel to combat the line-of-sight (LoS) blockage and to enhance the corresponding achievable data rate. Some recent efforts have investigated the integration of liquid crystal (LC)-RIS in the VLC receiver to also improve the corresponding achievable data rate. To jointly benefit from the previously mentioned appealing capabilities of the RIS technology in both the channel and the receiver, in this work, we propose a novel indoor VLC system that is jointly assisted by a mirror array-based RIS in the channel and an LC-based RIS aided-VLC receiver. To illustrate the performance of the proposed system, a rate maximization problem is formulated, solved, and evaluated. This maximization problem jointly optimizes the roll and yaw angles of the mirror array-based RIS as well as the refractive index of the LC-based RIS VLC receiver. Moreover, this maximization problem considers practical assumptions, such as the presence of non-users blockers in the LoS path between the transmitter-receiver pair and the user's random device orientation (i.e., the user's self-blockage). Due to the non-convexity of the formulated optimization problem, a low-complexity algorithm is utilized to get the global optimal solution. A multi-user scenario of the proposed scheme is also presented. Furthermore, the energy efficiency of the proposed system is also investigated. Simulation results are provided, confirming that the proposed system yields a noteworthy improvement in data rate and energy efficiency performances compared to several baseline schemes.
\end{abstract}

\begin{IEEEkeywords}
 Reflecting intelligent surface (RIS), visible light communication (VLC), liquid crystals (LCs), random receiver orientation, achievable
 rate, energy efficiency (EE), line of sight (LoS) blockage.
\end{IEEEkeywords}

\vspace{-0.8em}
\section{Introduction}

\IEEEPARstart{T}{he} exponential increase in connected devices and the ongoing development of wireless applications are driving the need to explore alternative wireless communications options to \gls{RF} communications. \Gls{VLC} has emerged as a bandwidth-abundant, cost-effective, and secure communications technology. Subsequently, \gls{VLC} is seen as a promising candidate for complementing \gls{RF} communications in future wireless networks~\cite{8528460}.

In ordinary VLC systems, the \gls{LoS} availability between the transmitter and the receiver is essential for reliable data transmission~\cite{9543660}. However, there may be other users (i.e., blockers) or obstacles that can obstruct this direct path, leading to \gls{LoS} blockage which is common in indoor \gls{VLC} systems~\cite{9226495}. The \gls{LoS} path can also be blocked when the device orientation of the intended user is not aligned with the transmitter (i.e., self-blockage). Previous studies on \gls{VLC} systems often assume that users' devices are facing upwards towards the ceiling (e.g.,~\cite{9226495,8862946,9411734}), but in reality, users typically hold their devices in different positions. Research has shown that random device orientations do affect the quality and existence of \gls{LoS} paths~\cite{8519786}, so taking this into account is important when designing and analyzing \gls{VLC} systems. Recently researchers have started unveiling the potential of integrating \glspl{RIS} on the walls to improve the communication links by reflecting waves from the transmitter toward the receiver. In case the \gls{LoS} path is blocked, an \gls{RIS} can reconfigure the wireless propagation channel to overcome this blockage. Accordingly, the adoption of \gls{RIS} can relax the \gls{LoS} requirement in \gls{VLC} systems.

On the other hand, an ordinary \gls{VLC} receiver typically includes a photo-detector (a \gls{PD}, a photo-transistor, etc.), and a convex lens. In \gls{VLC} systems, the incoming light must fall within the photo-detector's \gls{FoV} to recover the transmitted data successfully. To achieve this, convex lenses are used in ordinary \gls{VLC} receivers as extent reducers to collect and focus the incoming light onto the photo-detector's active area. However, the use of a convex lens can lead to a reduction in the amount of incident light power of up to 30\% because of reflection occurring at its top surface~\cite{ndjiongue2021re}. Additionally, convex lenses are not capable of dynamically steering the incoming light beam, which can limit the receiver's detection capabilities, especially with a large angle of incidence~\cite{9910023}. To overcome the aforementioned shortcoming of ordinary \gls{VLC} receivers, Ndjiongue \textit{et al.}~\cite{ndjiongue2021re} have proposed using voltage-controlled tunable \gls{LC} in the receiver to adjust the direction of the incident light and to provide intensity amplification. Subsequently, this can enhance the strength of the received signal and increase the corresponding achievable data rate. 

When it comes to the adoption of \gls{RIS} in \gls{VLC} systems, two main streams have evolved over the past few years, namely, (i) integrating a meta surface-based \gls{RIS}~\cite{sun2023optical,9348585,salehiyan2022performance,10008547,9838853,10047999} or mirror array-based \gls{RIS}~\cite{9526581,9543660,9500409,abumarshoud2022intelligent,9714890,10024150,wu2022configuring,shi2022performance,yang2022average,9784887} on walls between the transmitter and the receiver (noting that the performance of the mirror array-based \gls{RIS} outperforms the meta surface-based \gls{RIS}~\cite{9276478}), and (ii) incorporating an \gls{LC}-based \gls{RIS} in the receiver~\cite{9860058, 9609592,9910023,wu2023asymptotic} for light amplification and beam steering. However, to the best of the authors' knowledge, no one has investigated the joint benefit of integrating mirror array-based \gls{RIS} on walls between the transmitter and the receiver and an \gls{LC}-based \gls{RIS}-aided receiver. Hence, this work intends to unveil the \gls{VLC} system performance gain under such a joint design. 

In this paper, an indoor \gls{VLC} system that jointly benefits from the \gls{RIS} technology in both the channel and the receiver is proposed to overcome the \gls{LoS} blockage problem that comes from both non-user blockers and self-blockage. The channel-assisted \gls{RIS} improves the reliability of the \gls{VLC} system, while the \gls{LC}-based \gls{RIS}-aided \gls{VLC} receiver, with its amplification and light steering capabilities, enhances the strength of the received signal. The main contributions can be summarized as follows:
\begin{itemize}
    \item A novel indoor \gls{VLC} system that is jointly assisted by a mirror array-based \gls{RIS} and an \gls{LC}-based \gls{RIS}-aided \gls{VLC} receiver to enhance the corresponding achievable data rate is proposed. The proposed system takes into consideration the effects of non-user blockers and user's device orientation.
    \item A rate maximization problem for the proposed system that consists of a joint design of the roll and yaw angles of the mirror array-based \gls{RIS} as well as the refractive index of the \gls{LC}-based \gls{RIS}-aided \gls{VLC} receiver is formulated. Because the formulated multi-variate optimization problem is non-convex, a sine-cosine-based optimization algorithm is employed to obtain the global optimal solution.
    \item An additional scenario that considers the wall reflection with no \gls{RIS} is analyzed and evaluated to illustrate the benefit of the integration of the \gls{RIS} technology in improving the performance of \gls{VLC} systems. Moreover, a multi-user scenario of the proposed scheme is presented, confirming that the proposed system yields a noteworthy improvement to VLC systems. Furthermore, an \gls{EE} maximization problem for the proposed system is analyzed and solved because of the importance of this metric in evaluating \gls{VLC} systems.
    \item Extensive simulation results are provided to illustrate the significant performance gains of the proposed joint system in terms of achievable data rate and \gls{EE} when compared with several baseline schemes. 
\end{itemize}
The structure of this paper is as follows. The system and channel models of the proposed \gls{RIS}-assisted \gls{VLC} system that jointly consider both channel-assisted \gls{RIS} and an \gls{LC}-based \gls{RIS}-aided receiver are presented in Section~\ref{Sec: System and Channel Models}. In Section~\ref{Sec: Achievable Data Rate Optimization}, the data rate optimization problem is formulated, and the sine-cosine-based optimization algorithm that gets the optimal solution for the formulated problem is presented. A scenario that considers the wall reflection with no \gls{RIS} is presented in Section~\ref{Sec: Wall reflection's Scenario}. In Section~\ref{Sec: Multi-user Scenario}, a multi-user scenario that utilizes the power domain non-orthogonal multiple access (NOMA) scheme is presented. The optimization of the energy efficiency metric (as an additional metric) for the proposed \gls{VLC} system is provided in Section~\ref{Sec: Energy Efficiency Optimization}. Extensive simulation results are provided in Section~\ref{Sec: Simulation Results}, which is followed by the paper's conclusion and future research directions in Section~\ref{Sec: Conclusion}.

\begin{figure*}[t!]
\centering
\vspace{-2em}
\includegraphics[width=0.8\textwidth]{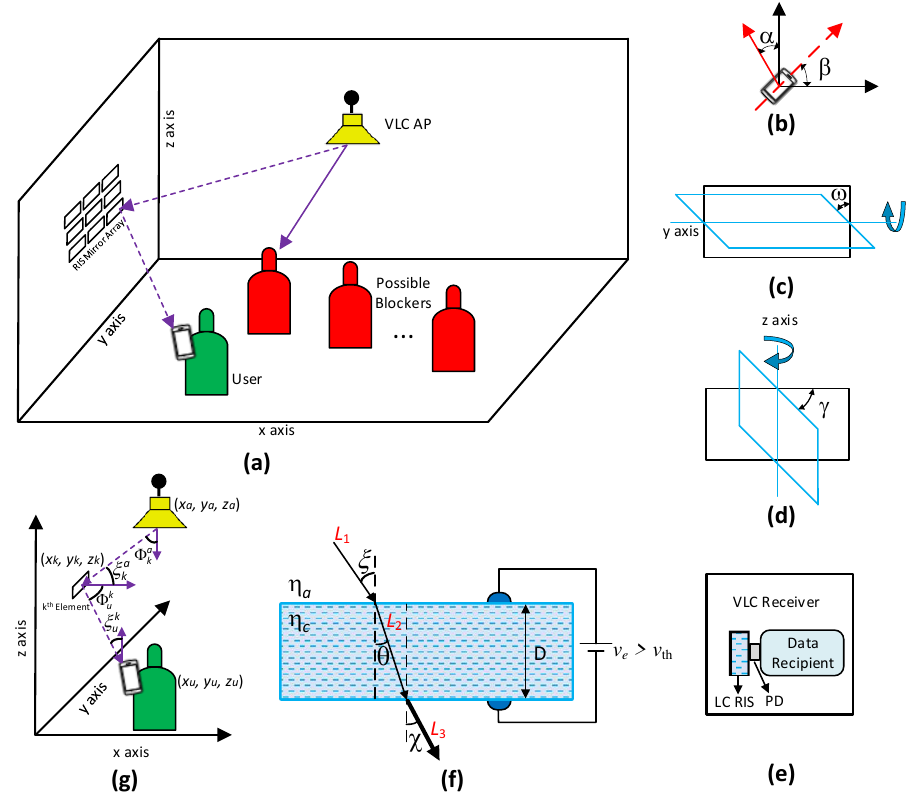}
\caption{Illustrations of the proposed joint mirror array and \gls{LC}-based \gls{RIS}-aided \gls{VLC} system with random device orientation. (a) The proposed system with one \gls{VLC} \gls{AP}, one user, a mirror array-based \gls{RIS}, and non-user possible blockers, (b) the orientation of the user device with respect to the device's azimuth angle $\beta$ and the device's polar angle $\alpha$, (c) the orientation of the mirror array-based \gls{RIS} with respect to the roll angle $\omega$, (d) the orientation of the mirror array-based \gls{RIS} with respect to the yaw angle $\gamma$, (e) a schematic of the \gls{LC}-based \gls{RIS}-aided receiver, (f) light signal propagation through the \gls{LC}-based \gls{RIS}-aided receiver, and (g) light signal propagation from the \gls{VLC} \gls{AP} (with a position vector $(x_a, y_a, z_a)$) to the intended user (with a position vector $(x_u, y_u, z_u)$) through the $k$-th element of the mirror array-based \gls{RIS} (with a position vector $(x_k, y_k, z_k)$), while specifying the angles of irradiance and incidence of the light signal.}
\label{fig: RIS-assisted VLC system model}
\vspace{-1.5em}
\end{figure*}

\vspace{-0.5em}
\section{System and Channel Models}
\label{Sec: System and Channel Models}
This section discusses the details of the indoor \gls{VLC} environment, including the \gls{VLC} system and channel models for the \gls{LoS} path and the \gls{NLoS} path resulting from the inclusion of the mirror array-based \gls{RIS} in the channel, then the structure of the \gls{LC}-based \gls{RIS}-aided receiver with its light steering and amplification capabilities.

\vspace{-0.5em}
\subsection{Indoor VLC Network}
We consider an indoor downlink \gls{VLC} system that is jointly assisted by a mirror array-based \gls{RIS} and an \gls{LC}-based \gls{RIS}-aided \gls{VLC} receiver as depicted in Fig.~\ref{fig: RIS-assisted VLC system model}. In Fig.~\ref{fig: RIS-assisted VLC system model}(a), the intended \gls{LoS} path is denoted by a solid line and the \gls{RIS}-assisted \gls{NLoS} path is denoted by a dotted line. Also, multiple non-user blockers are randomly deployed in the system. If any of these blockers lies in the way of the \gls{LoS} path, then the user would suffer from a signal outage. A signal outage might also occur if the user's device orientation is not perfectly aligned with the transmitter (i.e., user self-blockage). Hence, to overcome the outage from non-user blockers, an alternative path using the mirror array-based \gls{RIS} is proposed. Additionally, to circumvent the outage from self-blockage, an \gls{LC}-based \gls{RIS}-aided receiver with light steering capability is proposed. 

The orientation of the user's device is shown in Fig.~\ref{fig: RIS-assisted VLC system model}(b), where the user can move his device in any direction, which can be determined by the device's azimuth angle, $\beta$, and the polar angle, $\alpha$. The mirror array-based \gls{RIS} is composed of several passive reflecting elements and deployed on a wall between the transmitter and the receiver. Fig.~\ref{fig: RIS-assisted VLC system model}(c) and Fig.~\ref{fig: RIS-assisted VLC system model}(d), show the orientation of an arbitrary element of the mirror array-based \gls{RIS} with respect to the roll angle, $\omega$, and the yaw angle, $\gamma$, respectively. Fig.~\ref{fig: RIS-assisted VLC system model}(e) presents a schematic of an \gls{LC}-based \gls{RIS}-aided receiver, where an \gls{LC}-based \gls{RIS} module is placed in front of the \gls{PD}. The \gls{LC}-based \gls{RIS} module is made up of a series of thin layers, within which tin oxide nano-disks that have \gls{LC} infiltration are located, forming what is known as the \gls{LC} cell. These layers include (i) an anti-reflection polarizer for filtering incoming light, (ii) a glass substrate to orient the \gls{LC} molecules in a preferred direction, (iii) indium tin oxide for managing heat production and control, and (iv) a photo-alignment film that guides the light beam through the \gls{LC} cell~\cite{ndjiongue2021re}. 

Fig.~\ref{fig: RIS-assisted VLC system model}(f) illustrates how light propagates within the \gls{LC} cell when an external voltage, $v_\textnormal{e}$, exceeding the threshold voltage, $v_\textnormal{th}$, is applied. The emitted light signal $L_1$ from the \gls{VLC} \gls{AP} travels through the air medium (from the mirror array-based \gls{RIS}) with a refractive index $\eta_a$ and reaches the interface between the air and the \gls{LC} cell at an angle $\xi$. Since no light absorption occurs at this interface, a portion of the light signal $L_1$ is reflected while the remaining signal, $L_2$, is refracted at an angle $\theta$ as it passes through the \gls{LC} cell, which has a thickness $D$ and refractive index $\eta_c$. The electric field-induced molecular reorientation controls the propagation characteristics (such as direction and intensity) of the light signal as it passes through and exits the \gls{LC} cell. This reorientation induces changes in the refractive index, $\eta_c$, enabling the \gls{LC} \gls{RIS} to control the wave-guiding capability primarily through the refractive index. Finally, Fig.~\ref{fig: RIS-assisted VLC system model}(g) provides the angles of irradiance and incidence of the light signal starting from the \gls{VLC} \gls{AP} through the mirror array-based \gls{RIS} to the intended user. The light signal irradiates from the \gls{VLC} \gls{AP} with an angle $\Phi_k^a$, hits the mirror array-based \gls{RIS} with an angle $\xi_k^a$, is reflected by the mirror array-based \gls{RIS} with an angle $\Phi_u^k$, and finally strikes the \gls{LC}-based \gls{RIS} surface with an angle $\xi_u^k$. 
\vspace{-1em}
\subsection{VLC Channel: Mirror Array-Based RIS}
In the proposed \gls{VLC} system, the channel model can be defined as the signal propagation through the air and the \gls{LC} cell. The \gls{DC} gain represents the signal propagation through the air, while the transition coefficient represents the signal propagation through the \gls{LC} cell. The gain of the channel between the \gls{AP} and the user is expressed as~\cite{9543660}
\begin{equation} \label{eq: Total Channel Gain}
H=\iota H_{\textnormal{LoS}} \times \psi_{\textnormal{LC-LoS}} + \sum_{k=1}^\mathcal{K} H_{\textnormal{NLoS}}^{\textnormal{RIS}_k} \times \psi_{\textnormal{LC-NLoS}}, 
\end{equation}
\noindent where $\iota \in \{0,1\}$ is an indicator function that represents whether or not the \gls{LoS} path is obstructed. $H_{\textnormal{LoS}}$ denotes the channel gain of the \gls{LoS} path, $H_{\textnormal{NLoS}}$ represents the \gls{NLoS} channel gain, and $\psi_{\textnormal{LC-LoS}}$ and $\psi_{\textnormal{LC-NLoS}}$ denote the transition coefficient of the LoS and NLoS paths, respectively, which are discussed in Section~\ref{subsection: LC-Based RIS-Aided Receiver: Amplification Gain and Light Steering}. $\mathcal{K}$ denotes the number of squared surfaces (i.e., elements) in the mirror array-based \gls{RIS}. One can determine whether or not there is a reliable signal from the LoS path by examining if the received signal falls above the sensitivity of the photo-detector at the receiver. If the signal's strength is above the sensitivity of the photo-detector then $I$ is set to one, otherwise, it is set to zero. For example, to achieve a bit-error-rate (BER) of less than $10^{-12}$, the sensitivity of the photo-detector should be $-35$ dBm or above~\cite{hui2022fiber}.

The channel gain for the \gls{LoS} path is expressed as~\cite{1277847}
\begin{equation} \label{eq: The channel gain of the LoS link}
    H_{\textnormal{LoS}}= \begin{cases} \frac{(m+1)A_{\textnormal{PD}}}{2\pi d^2}\cos^m(\Phi) \cos(\xi) G(\xi) T(\xi), \\
    \ \ \qquad \qquad 0 \leq \xi \leq \xi_{\textnormal{FoV}} \\
    0, \qquad \qquad \xi > \xi_{\textnormal{FoV}}, \end{cases}
\end{equation}
\noindent where $m$ denotes the Lambertian index and is equal to $\frac{-1}{\textnormal{log}_2(\cos (\Phi_{1/2}))}$, with $\Phi_{1/2}$ as the semi-angle of the \gls{VLC} \gls{AP}, $A_{\textnormal{PD}}$ denotes the physical area of the \gls{PD}, $d$ denotes the distance between the \gls{VLC} \gls{AP} and the intended user, $\Phi$ and $\xi$ denote the angle of irradiance and angle of incidence, respectively, $T(\xi)$ denotes the gain of the optical filter, $G(\xi)$ denotes the gain of the optical concentrator, and $\xi_{\textnormal{FoV}} \leq \frac{\pi}{2}$ denotes the \gls{PD}'s \gls{FoV}. The gain of the optical concentrator is $G(\xi)=f^2 / \sin^2 \xi_{\textnormal{FoV}}$, $0 \leq \xi \leq \xi_{\textnormal{FoV}}$, where $f$ denotes the refractive index. The device's orientation highly influences the value of $\xi$, which can be expressed by both the azimuth and polar angles of the device as follows~\cite{8540452}
\begin{equation} \label{eq: Cos Xi}
\begin{split}
\cos(\xi) = & \big(\frac{x_a - x_u}{d}\big) \cos(\beta) \sin(\alpha) + \\ & \big(\frac{y_a - y_u}{d}\big) \sin(\beta) \sin(\alpha) + \\
 &  \big(\frac{z_a - z_u}{d}\big) \cos(\alpha), 
\end{split}
\end{equation}
\noindent where $(x_a, y_a, z_a)$ denotes the position vector of the \gls{VLC} \gls{AP}, and $(x_u, y_u, z_u)$ the position vector of the user. The mirror array-based \gls{RIS} is divided into $\mathcal{K}$ squared surfaces (i.e., elements), and each element has an area $dA$. In this paper, similar to~\cite{9276478,9910023,9543660}, we assume that the incoming optical signal from the \gls{AP} hits the middle of the reflecting surfaces. The channel gain for the reflected signal from the $k$-th mirror array is obtained by~\cite{9276478}
\begin{equation} \label{eq: The channel gain of the NLoS link - RIS}
    H_{\textnormal{NLoS}}^{\textnormal{RIS}_k}(\gamma,\omega)= \begin{cases} \rho_{\textnormal{RIS}}\frac{(m+1)A_{\textnormal{PD}}}{2\pi^2 (d_k^a)^2 (d_u^k)^2} {dA}_k \cos^m(\Phi_k^a) \cos(\xi_k^a) \\
    \quad \times \ \cos(\Phi_u^k) \cos(\xi_u^k) G(\xi) T(\xi), \\ \quad \ 0 \leq \xi_u^k \leq \xi_{\textnormal{FoV}} \\
    0, \ \xi_u^k > \xi_{\textnormal{FoV}}, \end{cases}
\end{equation}
where $\rho_{\textnormal{RIS}}$ denotes the reflection coefficient of an \gls{RIS} element, $d_k^a$ represents the distance between the \gls{AP} and the $k$-th reflecting element, $d_u^k$ represents the distance between the $k$-th reflecting element and the intended user, ${dA}_k$ is a reflective area of a small region, $\Phi_k^a$ is the irradiance angle from the \gls{AP} toward the $k$-th reflecting element, $\xi_k^a$ represents the incidence angle on the $k$-th reflecting element, $\Phi_u^k$ represents the irradiance angle from the $k$-th reflecting element towards the intended user, and $\xi_u^k$ represents the incidence angle of the reflected signal at the user. The value of $\cos(\xi_u^k)$ can be obtained using~\eqref{eq: Cos Xi}, while the value of $\cos(\Phi_u^k)$ is given by~\cite{9910023}
\begin{equation} \label{eq: Cos Phi}
\begin{split}
\cos(\Phi_u^k)= & \big(\frac{x_k - x_u}{d_k^u}\big) \sin(\gamma) \cos(\omega) + \\
& \big(\frac{y_k - y_u}{d_k^u}\big) \cos(\gamma) \cos(\omega) + \\
& \big(\frac{z_k - z_u}{d_k^u}\big) \sin(\omega), 
\end{split}
\end{equation}
\noindent where $(x_k, y_k, z_k)$ denotes the position vector of the $k$-th reflecting element of the mirror array-based \gls{RIS}.
\vspace{-1em}
\subsection{LC-Based RIS-Aided Receiver: Amplification Gain and Light Steering}
\label{subsection: LC-Based RIS-Aided Receiver: Amplification Gain and Light Steering}
This subsection explains how the \gls{LC}-based \gls{RIS} module can enhance signal reception by amplifying and steering the incoming light, as shown in Fig.~\ref{fig: RIS-assisted VLC system model}(f). This figure shows that the light intensity from the \gls{LC} module at the refraction angle $\chi$ (labeled as $L_3$) is greater than the intensity of the incoming light, $L_1$. This light amplification occurs through stimulated emission, where photons from the incoming light interact with excited molecules in the \gls{LC} module that have been stimulated by an external voltage, resulting in the creation of new, coherent photons. $L_3$ can be calculated using Beer's absorption law~\cite{demtroder2014laser} and is given by
\begin{equation} \label{eq: The intensity of the output beam}
L3 = L1 \times \textnormal{exp}(\Gamma D) \times \psi_{\textnormal{LC}}, 
\end{equation}
where $\textnormal{exp}(\Gamma D)$ represents the exponential increase in the intensity of the incident light, $\psi_{\textnormal{LC}}$ denotes the transition coefficient, and $\Gamma$ symbolizes the amplification gain coefficient, which can be represented as~\cite{marinova1999photorefractive}
\begin{equation} \label{eq: Amplification gain coefficient}
\Gamma = \frac{2 \pi \eta_c^3 }{\lambda \cos(\xi_u^k)} r_{\textnormal{eff}} E, 
\end{equation}
where $\lambda$ denotes the wavelength of the optical signal, $E$, measured in [\textit{V}/m], denotes the strength of the externally applied electric field, and $r_{\textnormal{eff}}$ denotes the electro-optic coefficient. From~\eqref{eq: Amplification gain coefficient}, it is evident that the amplification gain of the \gls{LC}-based \gls{RIS} is affected by the wavelength of the optical signal, the applied voltage, and the refractive index of the \gls{LC} module. Therefore, it is crucial to choose these values thoughtfully to achieve the best possible performance of the \gls{LC}-based \gls{RIS} module.

The transition coefficient, $\psi_{\textnormal{LC}}$, measures the effect of the \gls{LC}-based \gls{RIS} module on the total channel gain. It can be calculated by studying how light moves through the module as it enters, passes through the \gls{LC} cell, and exits it. The transition coefficient of the NLoS path (i.e., the mirror array-based RIS path) can be expressed as follows~\cite{9910023} 
\begin{equation} \label{eq: Transition Coefficient}
\begin{split}
\psi_{\textnormal{LC-NLoS}} &= T_{\textnormal{ac}}(\xi_u^k) \times T_{\textnormal{ca}}(\theta) \\
&= (1-R_{\textnormal{ac}}(\xi_u^k)) \times (1-R_{\textnormal{ca}}(\theta)),
\end{split}
\end{equation}
where $T_{\textnormal{ac}}(\xi_u^k)$ denotes the angular transmittance of the incident light on the interface between air and the \gls{LC} cell and indicates how much of the incident light is refracted through it, $T_{\textnormal{ca}}(\theta)$ denotes the angular transmittance as the light signal exits the \gls{LC} cell. Since no light is absorbed in both the interface between air and the \gls{LC} cell and at the interface between the \gls{LC} cell and air (i.e., when the light signal enters and exits the \gls{LC} cell, respectively), the light signal either gets reflected or refracted at these two interfaces. In \eqref{eq: Transition Coefficient}, $R_{\textnormal{ac}}(\xi_u^k)$ and $R_{\textnormal{ca}}(\theta)$ represent the amount of light that gets reflected when the light signal enters or exits the \gls{LC} cell, respectively. They can be expressed as follows~\cite{9910023}  
\begin{equation} \label{eq: angular reflectance ca}
\begin{split}
R_{\textnormal{ac}}(\xi_u^k) = & \frac{1}{2} \Bigg(\frac{\eta^2 \cos(\xi_u^k) - \sqrt{\eta^2 - \sin^2(\xi_u^k)}}{\eta^2 \cos(\xi_u^k) + \sqrt{\eta^2 - \sin^2(\xi_u^k)}}\Bigg)^2 + \\
&  \frac{1}{2} \Bigg(\frac{ \cos(\xi_u^k) - \sqrt{\eta^2 - \sin^2(\xi_u^k)}}{ \cos(\xi_u^k) + \sqrt{\eta^2 - \sin^2(\xi_u^k)}}\Bigg)^2,
\end{split}
\end{equation}
\begin{equation} \label{eq: angular reflectance ac}
\begin{split}
R_{\textnormal{ca}}(\theta) = & \frac{1}{2} \Bigg(\frac{\eta_1^2 \cos(\theta) - \sqrt{\eta_1^2 - \sin^2(\theta)}}{\eta_1^2 \cos(\theta) + \sqrt{\eta_1^2 - \sin^2(\theta)}}\Bigg)^2 + \\
&  \frac{1}{2} \Bigg(\frac{ \cos(\theta) - \sqrt{\eta_1^2 - \sin^2(\theta)}}{ \cos(\theta) + \sqrt{\eta_1^2 - \sin^2(\theta)}}\Bigg)^2,
\end{split}
\end{equation}
where $\eta = \eta_c/\eta_a$ and $\eta_1 = \eta_a/\eta_c$ denote the relative refractive indices when the light enters and exists the \gls{LC}-based \gls{RIS}, respectively, $\eta_a$ and $\eta_c$ are the refractive indices of the air medium and the \gls{LC} cell, respectively. If the LoS path exists,~\eqref{eq: Transition Coefficient} becomes $\psi_{\textnormal{LC-LoS}} = T_{\textnormal{ac}}(\xi_u^a) \times T_{\textnormal{ca}}(\theta)$, where $\xi_u^a$ represents the angle of incidence from the VLC AP on the LC-based RIS surface.

By examining~\eqref{eq: Transition Coefficient} to~\eqref{eq: angular reflectance ac}, we can see that the transition coefficient can be optimized by adjusting the refractive index $\eta_c$ of the \gls{LC}-based \gls{RIS}. This involves modifying the tilt angle $\phi$, which determines the molecular orientation of the \gls{LC} cell. The relation between the tilt angle and the refractive index is given by~\cite{saleh2019fundamentals}
\begin{equation} \label{eq: Relationship between the refractive index and the tilt angle}
\frac{1}{\eta_c^2 (\phi)} = \frac{\cos^2(\phi)}{\eta_e^2} + \frac{\sin^2(\phi)}{\eta_o^2}, 
\end{equation}
where $\eta_c (\phi)$ represents the refractive index of the \gls{LC} cell at the specific tilt angle $\phi$, $\eta_e$ and $\eta_o$ correspond to the extraordinary and ordinary refractive indices of the \gls{LC} cell, respectively. It is worth noting that the tilt angle, $\phi$, is controlled by an externally applied voltage, and the relationship between the two can be described as~\cite{9910023} 
\begin{equation} \label{eq: tilt angle is controlled by an externally applied voltage}
    \phi= \begin{cases} 
    \frac{\pi}{2} - 2 \tan^{-1} \Bigg[\textnormal{exp}\Bigg(- \frac{v_{\textnormal{e}} - v_\textnormal{th}}{v_0}\Bigg)\Bigg], v_{\textnormal{e}} > v_{\textnormal{th}} \\
    0, \qquad \qquad \qquad \qquad \qquad \qquad \ \ v_{\textnormal{e}} \leq v_{\textnormal{th}}
 \end{cases}
\end{equation}
where $v_e$ refers to the externally applied voltage, $v_{\textnormal{th}}$ represents the critical voltage needed to initiate the tilting process, and $v_0$ is a fixed value. As the tilt angle, $\phi$, is controlled by the voltage applied to the \gls{LC} cell, according to~\eqref{eq: tilt angle is controlled by an externally applied voltage}, the \gls{LC} cell can be utilized as a voltage-controlled \gls{RIS}. This makes it possible to control the propagation of light by changing the refractive index, $\eta_c$, and refraction angle, $\phi$, to steer the incoming light signal.
\vspace{-1em}
\section{Achievable Data Rate Optimization}
\label{Sec: Achievable Data Rate Optimization}
In this section, we present the details of the achievable data rate maximization problem investigated for the proposed joint mirror array and \gls{LC}-based \gls{RIS}-aided \gls{VLC} system. Then, due to the non-convexity of the formulated multi-variate optimization problem, a sine-cosine-based optimization algorithm~\cite{mirjalili2016sca} is utilized to get the global optimal solution.
\vspace{-1em}
\subsection{Achievable Data Rate Maximization Problem}
The achievable data rate of the proposed system can be expressed by the following lower bound~\cite{6636053}
\begin{equation} \label{eq: Achievable Data Rate}
R_{\textnormal{VLC}} = B \textnormal{log}_2 \Bigg(1 + \frac{\textnormal{exp(1)}}{2 \pi} \frac{\Big(\frac{p}{q} R_{\textnormal{PD}} \textnormal{exp}(\Gamma D)  H  \Big)^2}{N_o B}  \Bigg),
\end{equation}
\noindent where $B$ denotes the system's bandwidth, $p$ denotes the optical power, $q$ represents the ratio of the electrical signal power to the optical transmit power, $R_{\textnormal{PD}}$ represents the responsivity of the \gls{PD}, and $N_o$ denotes the power spectral density of the noise. In order to maximize the rate, this paper proposes (i) controlling the orientation of the mirror array by its roll angle, $\omega$, and the yaw angle, $\gamma$, and (ii) adjusting the refractive index, $\eta_c$, of the \gls{LC}-based module. With this, the achievable data rate maximization problem can be formulated as
\begin{alignat}{3}
\textnormal{(P0):} &\underset{\{\omega,\gamma,\eta_c\}}{\text{max}} &\ & R_{\textnormal{VLC}}, \label{eq:objective1}\\
&\quad \ \text{s.t.} &  & \textnormal{C1:} - \frac{\pi}{2} \leq \omega \leq \frac{\pi}{2}, \label{eq:constraint-a} \\
&  &  & \textnormal{C2:} - \frac{\pi}{2} \leq \gamma \leq \frac{\pi}{2}, \label{eq:constraint-b}\\ 
&  &  & \textnormal{C3:} \quad 1.5 \leq \eta_c \leq 1.7, \label{eq:constraint-c} 
\end{alignat}
where~\eqref{eq:constraint-a} and~\eqref{eq:constraint-b} represent the bounds of the roll angle and yaw angle, respectively. Constraint~\eqref{eq:constraint-c} represents the bounds of a typical off-the-shelf LC E7 (Merck)~\cite{li2005refractive}. The above optimization problem is highly non-convex and cannot be solved using traditional optimization methods. Therefore, we utilize a metaheuristic approach based on the \gls{SCA}~\cite{mirjalili2016sca} to find the appropriate orientation angles for the mirror array as well as the appropriate refractive index for the \gls{LC} cell to achieve the best data rate performance. The \gls{SCA} method was chosen over other metaheuristics due to its benefits, namely, its (i) ease of implementation, (ii) fast convergence, and (iii) ability to avoid local optimum solutions.
\vspace{-2em}
\subsection{Proposed Solution Methodology}
\label{subsection: Proposed Solution Approach}
The \gls{SCA} is a population-based stochastic optimization method introduced by Mirjalili in~\cite{mirjalili2016sca}. The method starts by generating random candidate solutions and gradually moves them toward the optimal solution using a mathematical model based on the sine and cosine functions. In the \gls{SCA} method, at the beginning of each iteration, $t$, a set of search agents are randomly positioned within the feasible solution space of the optimization problem (P0). Each search agent represents a possible solution to the problem and its fitness is evaluated using the objective function~\eqref{eq:objective1}. The agent with the highest fitness is selected as the destination point $\mathcal{D}^t$. In the next iteration $t+1$, each agent updates its solution using the following formula~\cite{mirjalili2016sca}
\begin{equation} \label{eq: current solution}
    s_{n,\upsilon}^{t+1}= \begin{cases} 
     s_{n,\upsilon}^{t} + r_1 \times \cos (r_2) \times |r_3 \mathcal{D}^t - s_{n,\upsilon}^{t} | \ \textnormal{if} \  r_4 \geq 0.5, \\
    s_{n,\upsilon}^{t} + r_1 \times \sin (r_2) \times |r_3 \mathcal{D}^t - s_{n,\upsilon}^{t} | \ \textnormal{if} \  r_4 < 0.5,
 \end{cases}
\end{equation}
where $s_{n,\upsilon}^{t}$ refers to the solution of the $n$-th agent at the $t$-th iteration, while $s_{n,\upsilon}^{t+1}$ represents the current updated solution, and the symbol $|.|$ means the absolute value of the expression enclosed in it. The parameters $r_1$, $r_2$, $r_3$, and $r_4$ can be expressed as follows. $r_1=a - t \frac{a}{T}$, where $T$ is a predefined maximum iteration number and $a$ is a constant value. Depending on the value of the randomly generated parameter $r_1$, the movement direction for the current agent could be towards (if $r_1$ $<$ 1) or away from (if $r_1$ $>$ 1) the destination point. The parameter $r_2$ controls how far the movement should be towards or away from the destination point and its value is randomly chosen in the interval $(0,2\pi)$. The parameter $r_3$ determines how much the destination point affects the distance between the current solution and the destination point and its value is randomly chosen through uniform random distribution in the interval $(0,2)$. Finally, the parameter $r_4$ that falls in the interval $(0,1)$ switches randomly between the cosine and sine components. In~\eqref{eq: current solution}, $r_1 \times \cos(r_2)$ and $r_1 \times \sin(r_2)$ enables both exploration and exploitation during the search process. In particular, when these two values are greater than 1 or less than -1, the algorithm conducts a global exploration search. Conversely, when these two values fall within the range of -1 to 1, the algorithm performs an exploitation search. The algorithm guides agents toward the best-known positions in the search space and updates their solutions until a maximum number of iterations (i.e., a termination criterion) is reached. A summary of the proposed solution methodology for the rate maximization problem in (P0) is provided in Algorithm~\ref{Algorithm: proposed solution}. This paper assumes that Algorithm~\ref{Algorithm: proposed solution} is implemented by a central control unit, linked to both the \gls{RIS} mirrors and the \gls{AP}, which is capable of acquiring the necessary downlink channel state information.

\begin{algorithm}[!t]
\footnotesize
\DontPrintSemicolon
\SetAlgoLined
\caption{Proposed solution methodology for the rate maximization problem in (P0).} \label{Algorithm: proposed solution}
\LinesNumbered
\KwIn{$N$, $T$, and $a$;}
\KwOut{The best solution $s_{n^*}= (\omega^*,\gamma^*,\eta_c^*)$, where $n^*$ denotes an arbitrary search agent and $(\omega^*,\gamma^*,\eta_c^*)$ denotes the global optimal solution of (P0);}
\textbf{First Stage}\;
Set $t=0$;\;
Initialize the search agents with a set of random solutions
$s_{n,\upsilon}^{t}$, $\forall n$;\;
Determine the fitness of each search agent using~\eqref{eq: Achievable Data Rate};\;
Store the solution of the fittest search agent in~$\mathcal{D}^t$;\;
\textbf{Second Stage}\;
Set $t=1$;\;
\While{Termination criterion is not reached}{
Get the value of the parameters: $r_1$,$r_2$, $r_3$, $r_4$;\;
\For{$n=1:N$}{
 \For{$\upsilon=1:3$}{ 
 Update the solutions $s_{n,\upsilon}^{t}$, $\forall n$ using~\eqref{eq: current solution};\;
}
}
Check for search agents that violate the constraints of (P0) and remove the violations to guarantee the solutions' feasibility;\;
Update the fitness of each search agent using~\eqref{eq: Achievable Data Rate};\;
Update~$\mathcal{D}^t$ if there is any better solution;\;
Update the iteration counter $t=t+1$;\;
}
return $s_{n^*}= (\omega^*,\gamma^*,\eta_c^*)$;\;
\end{algorithm}
\vspace{-2em}
\subsection{Computational Complexity Analysis}
\label{Subsection: Computational Complexity Analysis}
The computational complexity of Algorithm~\ref{Algorithm: proposed solution} can be evaluated as follows. At first, the generation of the initial set of solutions for all agents necessitates $\mathcal{O}(NV)$ operations, where $N$ and $V$ are the numbers of search agents and decision variables, respectively. The process of evaluating the solution fitness for all agents needs $\mathcal{O}(N)$ operations, and selecting the destination point has a complexity of $\mathcal{O}(N)$. As a result, the first-stage computational complexity of Algorithm~\ref{Algorithm: proposed solution} is $\mathcal{O}(NV)$. The worst-case complexity for updating the solution sets based on~\eqref{eq: current solution} is $\mathcal{O}(NVT)$. The worst-case complexity for evaluating the updated solution fitness for all agents is $\mathcal{O}(NT)$, and is $\mathcal{O}(NT)$ for updating the destination point. Thus, the second-stage worst-case computational complexity of Algorithm~\ref{Algorithm: proposed solution} is $\mathcal{O}(NVT)$. Therefore, based on the aforementioned discussion, the overall worst-case computational complexity for finding a solution using Algorithm~\ref{Algorithm: proposed solution} is approximately $\mathcal{O}(NVT)$, which consists of the $\mathcal{O}(NV)$ complexity of the first stage and the $\mathcal{O}(NVT)$ complexity of the second stage.
\vspace{-1em}
\section{Wall reflection's Scenario}
\label{Sec: Wall reflection's Scenario}
In this section, we examine the scenario with no \gls{RIS}, but where the wall's reflection paths are considered. We only take into account first-order reflection since, as demonstrated in~\cite{9226495}, the performance of \gls{VLC} systems is minimally affected by higher-order reflections. The channel gain of the first-order reflection from the wall surface is expressed as~\cite{1277847}
\begin{equation} \label{eq: The channel gain of the NLoS link - Wall}
    H_{\textnormal{NLoS}}^{\textnormal{wall}}= \begin{cases} \rho_{\textnormal{wall}}\frac{(m+1)A_{\textnormal{PD}}}{2\pi^2(d_w^a)^2 (d_u^w)^2} {dA}_k \cos^m(\Phi_w^a) \cos(\xi_w^a) \cos(\Phi_u^w) \\
    \quad \times \ \cos(\xi_u^w) G(\xi) T(\xi), \ 0 \leq \xi_u^w \leq \xi_{\textnormal{FoV}} \\
    0, \qquad \qquad \qquad \qquad \quad \ \xi_u^w > \xi_{\textnormal{FoV}} \end{cases}
\end{equation}
where $\rho_{\textnormal{wall}}$ denotes the wall's surface reflection coefficient. 

The achievable rate of this proposed scenario can be expressed by the lower bound~\cite{6636053}
\begin{equation}
\begin{split} \label{eq: Achievable Data Rate wall}
& R_{\textnormal{VLC}}^{\textnormal{wall}} = B \textnormal{log}_2 \Bigg(1 + \frac{\textnormal{exp(1)}}{2 \pi} \times \\ & \resizebox{0.485\textwidth}{!}{$  \frac{\Big(\frac{p}{q} R_{\textnormal{PD}} \textnormal{exp}(\Gamma D)  \big( \iota H_{\textnormal{LoS}} \psi_{\textnormal{LC-LoS}} + H_{\textnormal{NLoS}}^{\textnormal{wall}} \psi_{\textnormal{LC-NLoS}}\big)  \Big)^2}{N_o B}  \Bigg). $}  
\end{split}
\end{equation}
\indent To maximize the achievable rate of this scenario, we replace the objective function of (P0) by~\eqref{eq: Achievable Data Rate wall}, and solve the resultant optimization problem by following the same steps mentioned in Section~\ref{subsection: Proposed Solution Approach}.
\vspace{-1em}
\section{Multi-user Scenario}
\label{Sec: Multi-user Scenario}
To implement this scenario, we assume that the number of intended users is $U$ and these users are sorted based on their channel gain $H_1 \leq H_2 \leq ... \leq H_U$~\cite{7792590}. Also, we utilize the power domain NOMA scheme to serve these users. According to the NOMA scheme, the VLC AP transmits a superposed signal, $x$, to serve the intended users, which can be represented as~\cite{shen2023secrecy}
\begin{equation} \label{eq: superposition coding}
    x = (\sum_{u=1}^{U} \sqrt{c_u P_\textnormal{S}} s_u) + I_\textnormal{DC},
\end{equation}
\noindent where $c_u$ and $s_u$ denote the allocated power ratio and the modulated message signal intended for the $u$-th user, respectively. $P_\textnormal{S}$ is the electrical transmit power of the signal, which is equal to $(\frac{p}{q})^2$. $I_\textnormal{DC}$ is a fixed bias current added to ensure the positive instantaneous intensity~\cite{7792590}. Based on~\cite{shen2023secrecy}, $c_u$ can be expressed as
\begin{equation} \label{eq: allocated power ratio}
c_u =  \begin{cases}
       \zeta (1-\zeta)^{u-1}, \ \textnormal{if} \ 1 \leq u<U \\
       (1-\zeta)^{u-1}, \quad \textnormal{if} \ u=U
       \end{cases}
\end{equation}
\noindent where $\zeta$ is a fixed value in the range $(0.5,1]$ and $\sum_{u=1}^U c_u~=~1$. The received signal at the $u$-th user, after removing the DC bias, is given by~\cite{shen2023secrecy}
\begin{equation} \label{eq: Received signal}
    y_u = H_u \times (\sum_{u=1}^{U} \sqrt{c_u P_\textnormal{S}} s_u) + z_u,
\end{equation}
\noindent where $z_u$ symbolizes the additive real-valued Gaussian noise with variance $\sigma^2$ (i.e., $z_u \sim \mathcal{N}(0,\sigma^2)$), which includes both the thermal and shot noises. According to the NOMA scheme, each user needs to perform successive interference cancellation (SIC) to decode its information~\cite{9154358}. For simplicity, we assume that there is no residual interference after performing the SIC process (i.e., perfect SIC)~\cite{maraqa2021achievable}. Accordingly, the sum rate of the proposed system can be expressed as
\begin{equation} \label{eq: Sum rate}
\begin{aligned}
& R_{\textnormal{sum}} =\sum_{u=1}^{U} R_u, \\
& \textnormal{where} \\
& R_u = \begin{cases} B \textnormal{log}_2 \Bigg(1 + \frac{\textnormal{exp(1)}}{2 \pi} \frac{\Big( R_{\textnormal{PD}} \textnormal{exp}(\Gamma D) H_u \Big)^2 c_u P_\textnormal{S}}{ I + N_o B}  \Bigg), 1 \leq u < U\\
B \textnormal{log}_2 \Bigg(1 + \frac{\textnormal{exp(1)}}{2 \pi}  \frac{\Big( R_{\textnormal{PD}} \textnormal{exp}(\Gamma D)  H_u  \Big)^2 c_u P_\textnormal{S}}{N_o B}  \Bigg), u=U
\end{cases}
\end{aligned}
\end{equation}
\noindent and $I=\sum_{i=u+1}^U \Big( R_{\textnormal{PD}} \textnormal{exp}(\Gamma D) H_u \Big)^2 c_i P_\textnormal{S}$ represents the inter-user interference term resulting from the application of the NOMA scheme. To maximize the sum rate metric of this scenario, we replace the objective function of (P0) by~\eqref{eq: Sum rate}, and solve the resultant optimization problem by following the same steps mentioned in Section~\ref{subsection: Proposed Solution Approach}.

\vspace{-1em}
\section{Energy Efficiency Optimization}
\label{Sec: Energy Efficiency Optimization}

The \gls{EE} has become a widely used performance metric in \gls{VLC} systems~\cite{9348585,9350598,zhan2022optimal, 8307185}. The \gls{EE} can be defined as the ``ratio between the \gls{VLC} system's achievable rate and the total consumed power" and is expressed in Bits/Joule~\cite{9350598}. The achievable rate of the proposed system is provided in~\eqref{eq: Achievable Data Rate}. The total consumed power of the proposed system includes, the power consumed at the transmitter, \gls{RIS}, and receiver. The consumed power at the transmitter mainly involves the power consumption of the signal power, digital to analog converter (DAC), filter, power amplifier, \gls{LED} driver, and transmitter external circuit. The consumed power at the mirror array-based \gls{RIS} only includes the power required to rotate all the elements of the mirror array, since the mirrors are passive elements. Lastly, the consumed power at the receiver mainly involves the power consumption of analog to digital converter (ADC), trans-impedance amplifier (TIA), filter, \gls{LC}, and receiver external circuit. Accordingly, the total power consumption of the proposed system is described by~\cite{8307185,zhan2022optimal}
\begin{equation}
\begin{aligned} \label{eq: total power consumption}
 & P_{\textnormal{total}} =  P_\textnormal{T} + P_\textnormal{RIS} + P_\textnormal{R},
 \\ & P_\textnormal{T} =  P_\textnormal{S} + P_\textnormal{DAC} + P_\textnormal{Filter} + P_\textnormal{PA} + P_\textnormal{Driver} + P_\textnormal{T-Circuit}, 
 \\ & P_\textnormal{RIS} =  P_\textnormal{m} \times \mathcal{K},  
 \\ & P_\textnormal{R} = P_\textnormal{ADC} + P_\textnormal{TIA} + P_\textnormal{Filter} + P_\textnormal{LC} + P_\textnormal{R-Circuit},
\end{aligned}
\end{equation}
\noindent where the power consumption parameters mentioned in~\eqref{eq: total power consumption} can be quoted from the data sheets of \gls{VLC} systems. According to the previous analysis, the \gls{EE} of the proposed system can be expressed as 
\begin{equation} \label{eq: EE}
    \textnormal{EE} = \frac{R_{\textnormal{VLC}}}{P_{\textnormal{total}}}.
\end{equation}
\indent To maximize the \gls{EE} metric, we replace the objective function of (P0) by~\eqref{eq: EE}, and solve the resultant optimization problem by following the same steps mentioned in Section~\ref{subsection: Proposed Solution Approach}.

\begin{table}[!t]
\centering
\vspace{-0.5em}
\caption{Simulation Parameters}
\vspace{-0.5em}
\label{tab: Simulation Parameters}
\resizebox{1.05\columnwidth}{!}{%
\begin{tabular}{|l|l||l|l|l}
\cline{1-4}
\textbf{Parameter} & \textbf{Value} & \textbf{Parameter} & \textbf{Value} &  \\ \cline{1-4}

$\Phi_{1/2}$ & $ 70^\circ$ & $A_{\textnormal{PD}}$ & $1.0 \ \textnormal{cm}^2$ &  \\ \cline{1-4}

$d$ & $2.5 \ \textnormal{m} $  & $T(\xi)$ & $1.0$ &  \\ \cline{1-4}
 
$\xi_{\textnormal{FoV}}$ & $85^{\circ}$ & $f$ & $1.5$ &  \\ \cline{1-4}
 
$\rho_\textnormal{wall}$ & $0.8$ & $\rho_\textnormal{RIS}$ & $0.95$ &  \\ \cline{1-4}
 
$\eta_a$ & $1.0$ & $\eta_e$ & $1.7$ &  \\ \cline{1-4}

$\eta_o$ & $1.5$ & $v_\textnormal{th}$ & $1.34$ V &  \\ \cline{1-4}

$v_0$ & $1.0$ V & $D$ & $0.75$ mm &  \\ \cline{1-4}

$\lambda$ & $\{510,670\}$ nm & $r_\textnormal{eff}$ & $12$ pm/V&  \\ \cline{1-4}

$B$ & $200$ MHz & $q$ & $3.0$ &  \\ \cline{1-4}
 
$R_\textnormal{PD}$ & $0.53$ A/W & $N_o$ & $10^{-21} \textnormal{A}^2/\textnormal{Hz}$ &  \\ \cline{1-4}

$N$ & $2$ & $V$ & $3$ &  \\ \cline{1-4}

$a$ & $2.0$ & $T$ & $400$ &  \\ \cline{1-4}

$P_\textnormal{DAC}$ & $175$ mWatt & $P_\textnormal{ADC}$ & $95$ mWatt &  \\ \cline{1-4}

$P_\textnormal{Filter}$ & $2.5$ mWatt & $P_\textnormal{Driver}$ & $2758$ mWatt &  \\ \cline{1-4}

$P_\textnormal{TIA}$ & $2500$ mWatt & $P_\textnormal{m}$ & $100$ mWatt &  \\ \cline{1-4}

$P_\textnormal{T-Circuit}$ & $3250$ mWatt & $P_\textnormal{R-Circuit}$ & $1.9$ mWatt &  \\ \cline{1-4}

$P_\textnormal{PA}$ & $280$ mWatt & $P_\textnormal{LC}$ & $320$ mWatt &  \\ \cline{1-4}

\end{tabular}%
}
\vspace{-2em}
\end{table}

\vspace{-1em}
\section{Simulation Results}
\label{Sec: Simulation Results}

In this section, we present detailed numerical results to evaluate the performance of the proposed joint mirror array and \gls{LC}-based \gls{RIS}-aided \gls{VLC} system. A list of the default parameters used in this section is provided in Table~\ref{tab: Simulation Parameters}. Beyond this list, based on the findings presented in~\cite{8540452}, the azimuth angle, $\beta$, follows a uniform distribution with a value ranging in the interval [$-\pi$,$\pi$]. On the other hand, the polar angle, $\alpha$, can be described by using the Laplace distribution with a mean of $41$ degrees and a standard deviation of $9$ degrees. The polar angle is typically within the range of [0,$\frac{\pi}{2}$]. The users in the system are represented as cylinders that have a radius of $0.15$ meters and a height of $1.65$ meters. The receiver is held by a user who is positioned at $0.85$ meters above the ground and $0.36$ meters away from the user's body. The mirror array-based \gls{RIS} consists of 10 rows and 30 columns of mirrors, with each mirror measuring $0.1$ meters by $0.1$ meters. Both the dimension of the considered mirror array-based \gls{RIS} (in the y-z plane) and the dimension of the wall of interest (in the x-z plane) is $1.0~\textnormal{m} \times 3.0~\textnormal{m}$. The room size is $5.0~\textnormal{m} \times 5.0~\textnormal{m} \times 3.0~\textnormal{m}$ and the \gls{VLC} \gls{AP} is positioned at the center of the room ceiling with a position vector $2.5~\textnormal{m} \times 2.5~\textnormal{m} \times 3.0~\textnormal{m}$. In the simulation of the NLoS scenario, we set $I$ to zero. All the parameters used in this section are quoted from~\cite{maraqa2021achievable, 9910023, 9543660, 8307185,zhan2022optimal,shen2023secrecy}.

\begin{figure}[t!]
\centering
\vspace{-1em}
\includegraphics[width=0.485\textwidth]{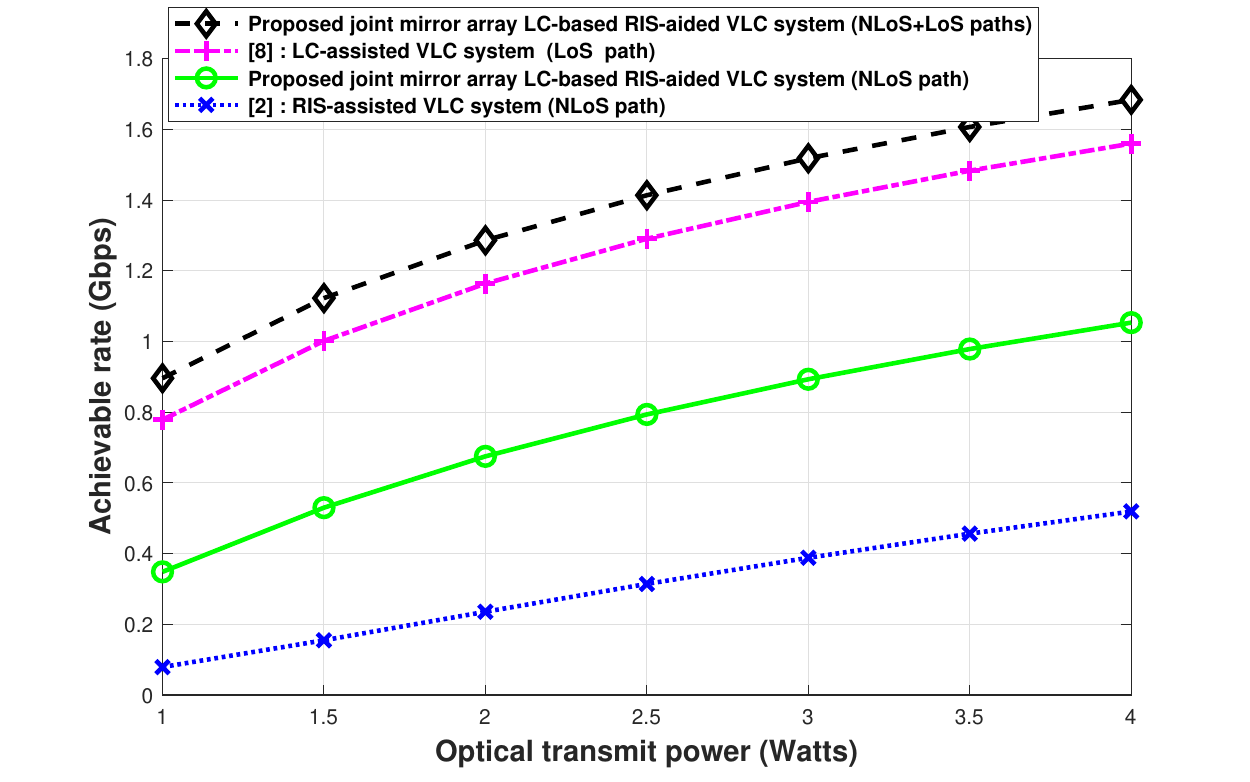}
\caption{The achievable data rate performance of the proposed joint mirror array \gls{LC}-based \gls{RIS}-aided \gls{VLC} system with both a NLoS-path case and a NLoS-plus-LoS-paths case compared to the following counterparts: (i) \gls{LC}-based \gls{RIS}-aided \gls{VLC} system~\cite{9910023} with a \gls{LoS} path, and (ii) \gls{RIS}-assisted \gls{VLC} system~\cite{9543660} with a \gls{NLoS} path. At $\lambda = 510$ nm. }
\label{fig: P_vs_R_current}
\vspace{-0.5em} 
\end{figure}

Fig.~\ref{fig: P_vs_R_current} compares the achievable data rate performance of the proposed joint mirror array \gls{LC}-based \gls{RIS}-aided \gls{VLC} system with both a NLoS-path case and a NLoS-plus-LoS-paths case to the following counterparts: (i) \gls{LC}-based \gls{RIS}-aided \gls{VLC} system with a \gls{LoS} path~\cite{9910023}, and (ii) \gls{RIS}-assisted \gls{VLC} system with a \gls{NLoS} path~\cite{9543660}. It can be observed that, when both the LoS and NLoS paths are considered, the proposed system achieves up to $115\%$ improvement in the data rate performance compared to the counterpart in (i). Also, when the LoS path is obstructed, the proposed system achieves up to $404\%$ improvement in the data rate performance compared to the counterpart in (ii). This illustrates that adopting the \gls{RIS} technology in both the channel and at the receiver contribute to a significant enhancement of the \gls{VLC} system's achievable data rate.

\begin{figure}[t!]
\centering
\includegraphics[width=0.485\textwidth]{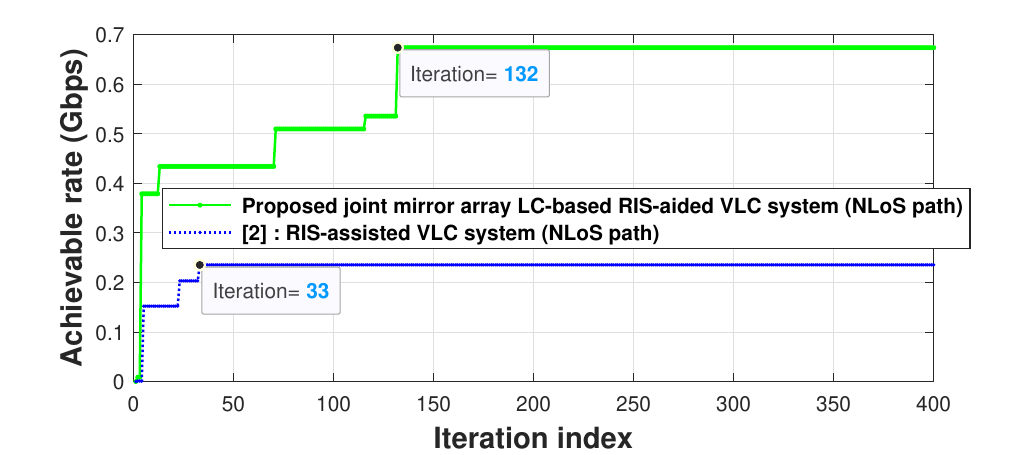}
\caption{The convergence curves of the \gls{SCA} algorithm for the proposed joint mirror array \gls{LC}-based \gls{RIS}-aided \gls{VLC} system compared to its counterpart of \gls{RIS}-assisted \gls{VLC} system~\cite{9543660}. At $p = 2$ Watts and $\lambda = 510$ nm. }
\label{fig: convergence curves}
\vspace{-1em}
\end{figure}

Fig.~\ref{fig: convergence curves} shows a convergence analysis of the \gls{SCA} algorithm for the proposed system compared to its counterpart of \gls{RIS}-assisted \gls{VLC} system. In this figure, the proposed joint mirror array \gls{LC}-based \gls{RIS}-aided \gls{VLC} system needs around $4$ times more iterations to converge to the global optimal solution, compared to its counterpart of \gls{RIS}-assisted \gls{VLC} system. This behavior is expected since the proposed system has a larger search space compared to its counterpart. Specifically, according to the analysis provided in Section~\ref{Subsection: Computational Complexity Analysis} and the simulation parameters in Table~\ref{tab: Simulation Parameters}, one can deduce that the search space (i.e., the overall worst-case computational complexity that is equal $\mathcal{O}(NVT)$) for the proposed system is double the search space of the \gls{RIS}-assisted counterpart. 

\begin{figure}[t!]
\centering
\includegraphics[width=0.485\textwidth]{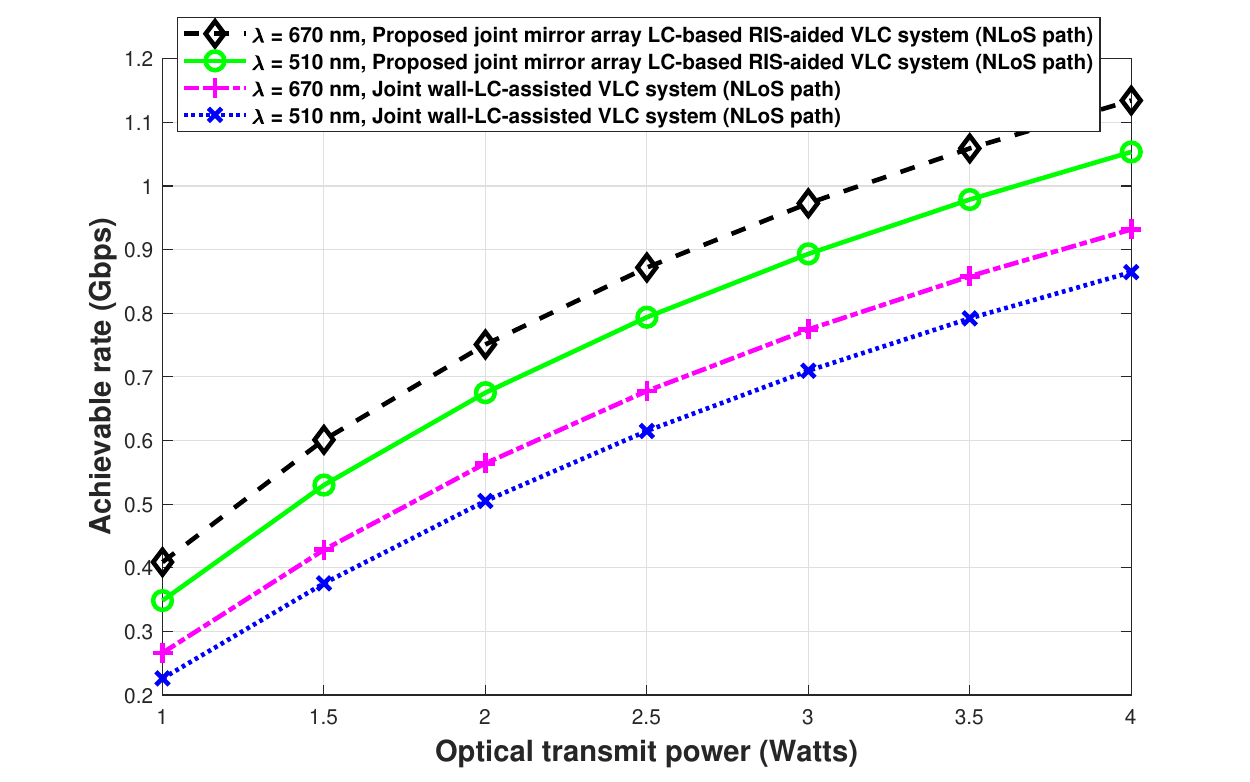}
\caption{The achievable data rate performance of the proposed joint mirror array \gls{LC}-based \gls{RIS}-aided \gls{VLC} system compared to its counterpart of wall-\gls{LC}-assisted \gls{VLC} system (mentioned in Section~\ref{Sec: Wall reflection's Scenario}) for light signals of different wavelengths.}
\label{fig: P_vs_R_wavelenght}
\vspace{-1em}
\end{figure}

\begin{figure}[t!]
\centering
\vspace{-0.5em}
\includegraphics[width=0.485\textwidth]{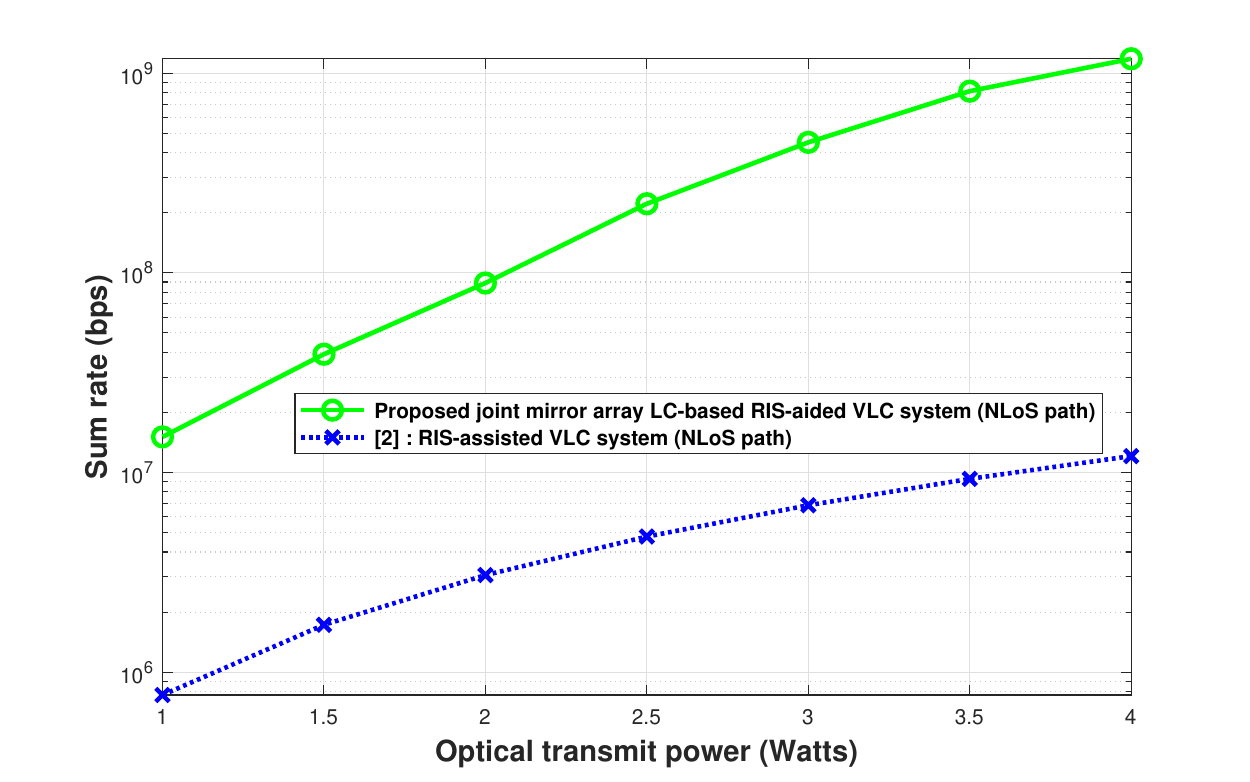}
\caption{The sum rate performance of the proposed joint mirror array LC-based RIS-aided VLC system compared to its counterpart of RIS-assisted VLC system~\cite{9543660}. At $\lambda = 510$ nm, $\zeta=0.6$, and $U=4$.}
\label{fig: SumR_multiuser_power}
\vspace{-1em}
\end{figure}

\begin{figure}[!t]
    \centering
    \vspace*{-0.2in}
    \subfloat[]{
        \hspace*{-2em}
        \includegraphics[scale=0.4]{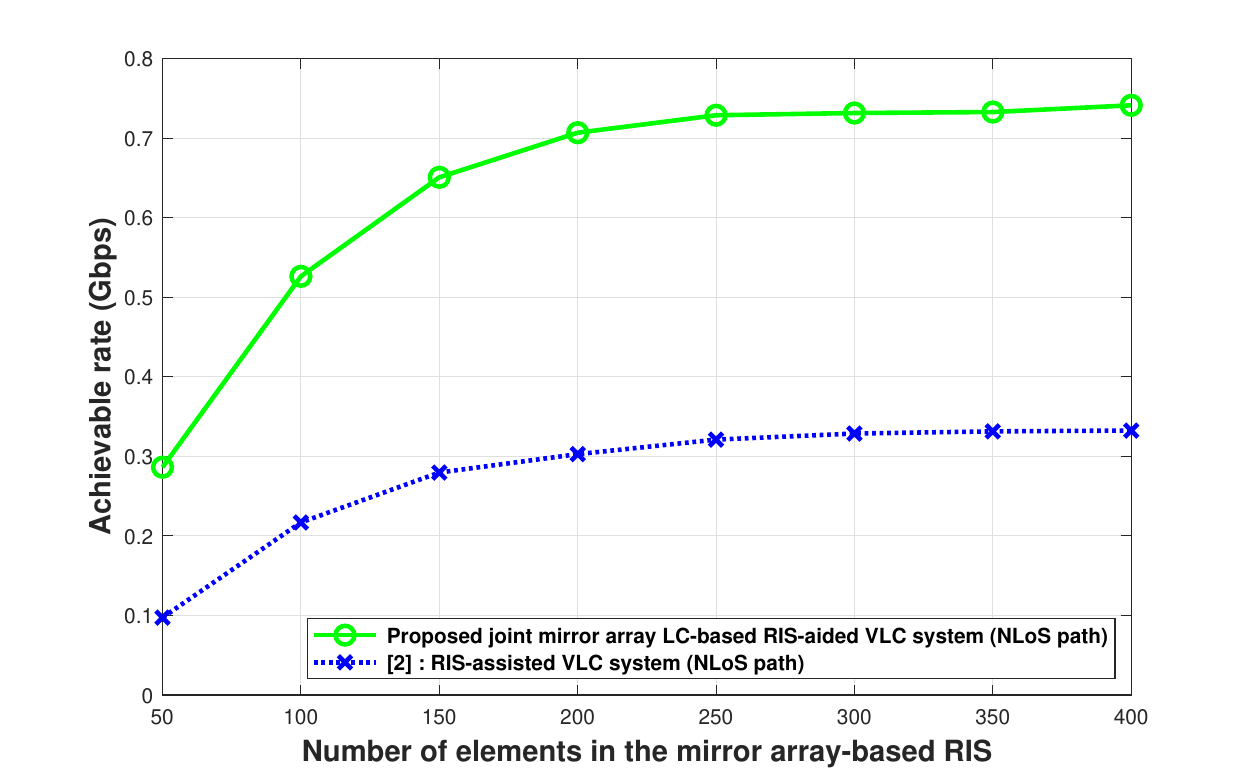}
        \label{fig: D_mirrors_vs_R}
    }
    \hfill
    \subfloat[]{
        \hspace*{-2em}
        \includegraphics[scale=0.4]{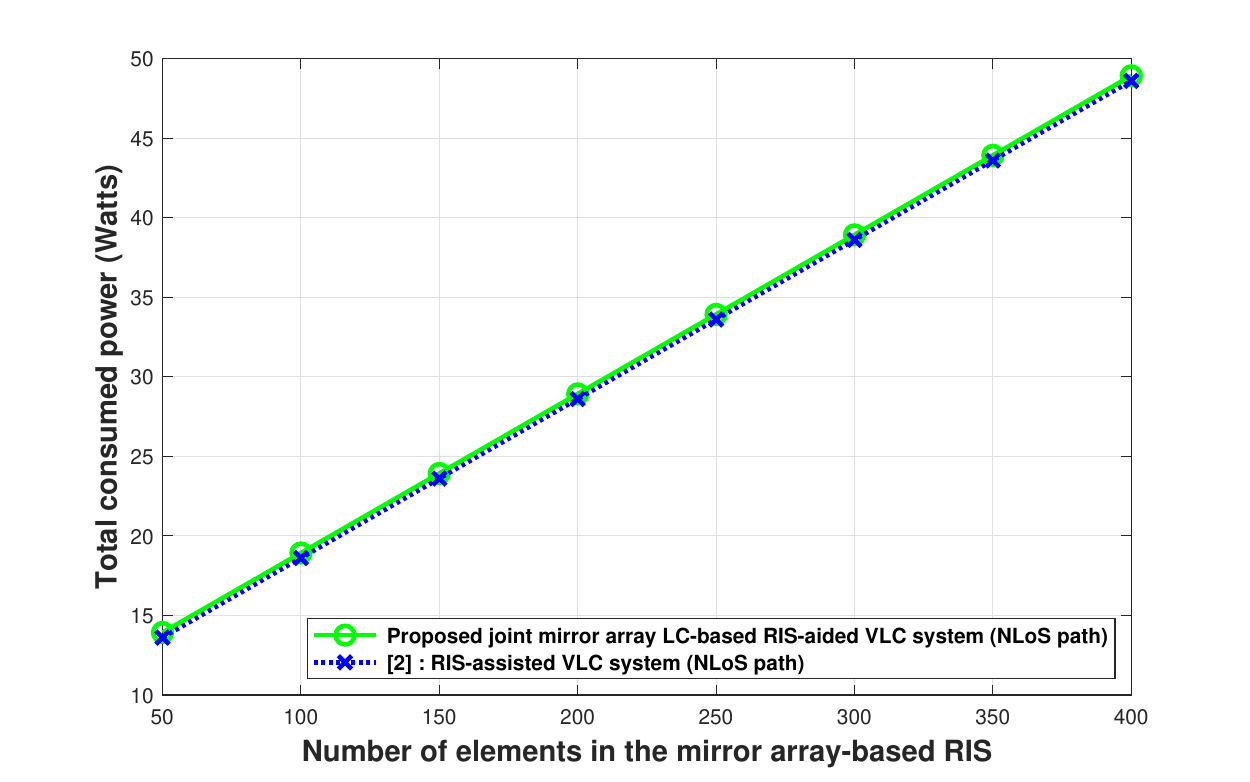}
        \label{fig: D_mirrors_vs_P}
    }
    \hfill
    \subfloat[]{
        \hspace*{-2em}
        \includegraphics[scale=0.4]{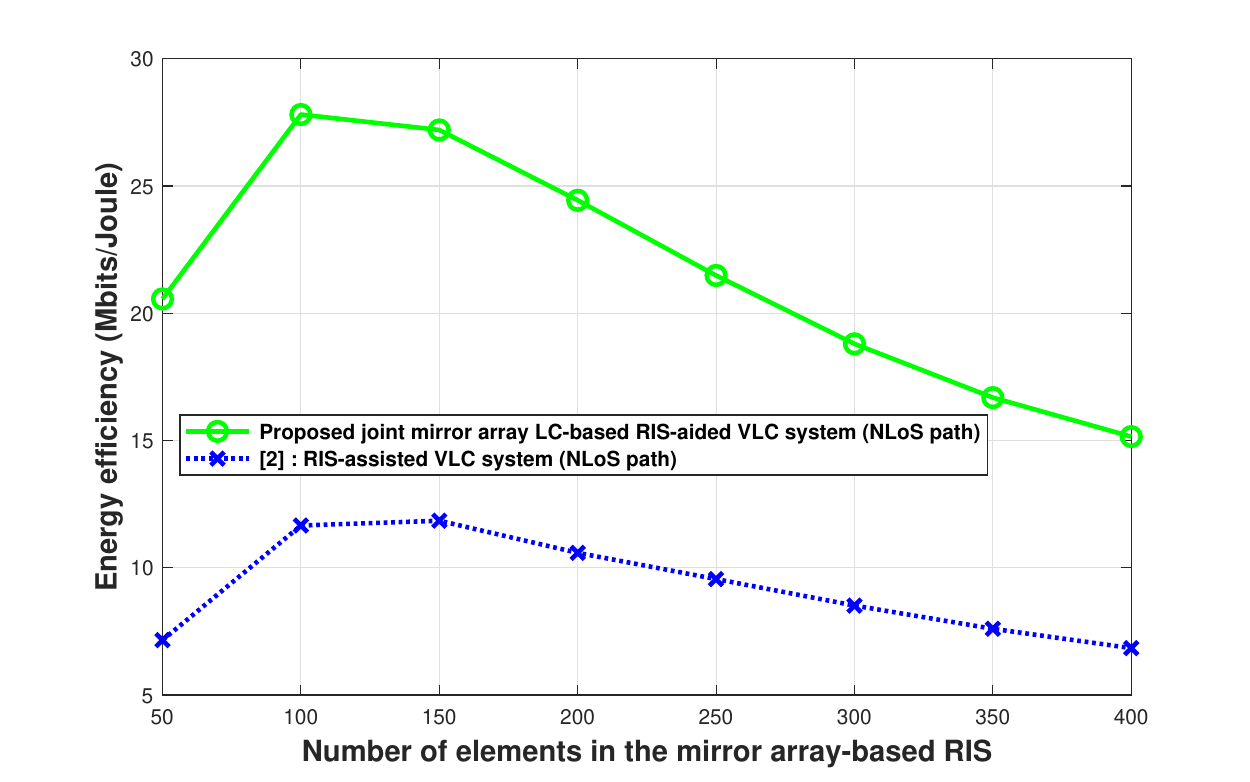}
        \label{fig: D_mirrors_vs_EE}
    }
    \caption{(a) The achievable data rate, (b) consumed-power, and (c) energy efficiency performance of the proposed joint mirror array \gls{LC}-based \gls{RIS}-aided \gls{VLC} system compared to its counterpart of \gls{RIS}-assisted \gls{VLC} system~\cite{9543660} for a different number of elements in the \gls{RIS} mirror array. At $p = 2$ Watts and $\lambda = 510$ nm. }
\label{fig: D_mirrors_vs_R_P_EE}
\vspace{-2em}
\end{figure}

Fig.~\ref{fig: P_vs_R_wavelenght} compares the achievable data rate performance of the proposed joint mirror array \gls{LC}-based \gls{RIS}-aided \gls{VLC} system to its counterpart of wall-\gls{LC}-assisted \gls{VLC} system (mentioned in Section~\ref{Sec: Wall reflection's Scenario}) for light signals of different wavelengths. This figure is simulated for different wavelengths since changing the wavelengths influences the data rate performance of the \gls{LC}-based \gls{RIS}-aided receiver. In this figure, changing the wavelength of the light signals from $510$ nm to $670$ nm achieves up to $117\%$ and $118\%$ improvement in the data rate performance for the proposed joint mirror array \gls{LC}-based \gls{RIS}-aided \gls{VLC} system and wall-\gls{LC}-assisted \gls{VLC} system, respectively. Additionally, the proposed joint mirror array \gls{LC}-based \gls{RIS}-aided \gls{VLC} system achieves up to $154\%$ improvement in the data rate performance compared to its counterpart of wall-\gls{LC}-assisted \gls{VLC} system. This demonstrates how pivotal the integration of the \gls{RIS} technology is in improving the performance of \gls{VLC} systems.

Fig.~\ref{fig: SumR_multiuser_power} compares the sum rate performance of the multi-user scenario (mentioned in Section~\ref{Sec: Multi-user Scenario}) for the proposed joint mirror array LC-based RIS-aided VLC system to its counterpart of RIS-assisted VLC system. In this figure, the provided sum rate results for the multi-user case shows a similar trend compared to the single intended user case (e.g., see Fig.~\ref{fig: P_vs_R_current}), where the proposed system achieves up to $196\%$ improvement in the data rate performance compared to the counterpart in~\cite{9543660}. Also, if we compare the sum rate performance of the multi-user case, in this figure, with the rate performance of the single intended user case reported in Fig.~\ref{fig: P_vs_R_current}, we can notice some rate performance degradation here. Such a trend is expected for two reasons, (i) in the adopted multi-user scenario, the NOMA scheme is utilized to serve the intended users simultaneously at the expense of introducing an inter-user interference that deteriorates the rate performance of the VLC system, and (ii) in the adopted multi-user scenario, the optical transmit power of the VLC AP is divided between the intended users (as explained in~\eqref{eq: allocated power ratio}).

Fig.~\ref{fig: D_mirrors_vs_R_P_EE} compares the achievable data rate, consumed-power, and energy efficiency performance of the proposed joint mirror array \gls{LC}-based \gls{RIS}-aided \gls{VLC} system compared to its counterpart of \gls{RIS}-assisted \gls{VLC} system~\cite{9543660} for a different number of elements in the mirror array-based \gls{RIS}. In Fig.~\ref{fig: D_mirrors_vs_R}, one can see that the achievable data rate performance of the proposed system increases steadily and then starts to saturate when the number of elements in the mirror array-based \gls{RIS} increases. This is because adding extra elements in a large mirror array-based \gls{RIS} provides marginal gain. A similar trend has been reported in~\cite{sun2023optical, wu2022configuring}. Fig.~\ref{fig: D_mirrors_vs_P} shows that the total power consumption of the proposed system is slightly higher than its counterpart of \gls{RIS}-assisted \gls{VLC} system. The extra power consumption comes from the inclusion of the \gls{LC}-based \gls{RIS} module in the \gls{VLC} receiver. From Fig.~\ref{fig: D_mirrors_vs_EE}, three main observations can be deduced, (i) as the number of elements in the mirror array-based \gls{RIS} increases the system's \gls{EE} increases then decreases. Such a trend in the system can be justified by the following argument: the increase in the number of elements is reflected as a logarithmic increase in the system's achievable data rate (Fig.~\ref{fig: D_mirrors_vs_R}) and as a linear increase in the total power consumption of the proposed system (Fig.~\ref{fig: D_mirrors_vs_P}). (ii) Although the proposed system consumes more power compared to its counterpart of \gls{RIS}-assisted \gls{VLC} system, it achieves up to $284\%$ improvement in the overall \gls{EE} performance. (iii) In our proposed system, the best \gls{EE} performance is obtained when the number of elements in the mirror array-based RIS is one hundred. At this point, the proposed \gls{VLC} system achieves a good trade-off between the achievable rate improvement and the total consumed power. Fig.~\ref{fig: D_mirrors_vs_EE} is obtained by replacing the objective function of (P0) by~\eqref{eq: EE} and then optimizing the \gls{EE} metric by following the same steps mentioned in Section~\ref{subsection: Proposed Solution Approach}.

\vspace{-0.5em}
\section{Conclusion and Future Research Directions}
\label{Sec: Conclusion}
In this paper, a novel indoor \gls{VLC} system that is jointly assisted by a mirror array-based \gls{RIS} and an \gls{LC}-based \gls{RIS}-aided \gls{VLC} receiver to enhance the corresponding achievable data rate is presented and investigated. The proposed system combats the \gls{LoS} blockage that results from (i) non-user blockers and (ii) the user's device orientation (i.e., self-blockage). A rate maximization problem and an \gls{EE} maximization problem are formulated, solved, and evaluated for the proposed system. These maximization problems jointly optimize the roll and yaw angles of the mirror array-based \gls{RIS} as well as the refractive index of the \gls{LC}-based \gls{RIS}-aided \gls{VLC} receiver. Because the formulated multi-variate optimization problems are non-convex, a sine-cosine-based optimization algorithm is employed to obtain the global optimal solution for both problems with very low complexity. Simulation results have revealed that adopting such joint mirror array \gls{LC}-based \gls{RIS}-aided deployment can dramatically improve the data rate as well as the \gls{EE} of indoor \gls{VLC} systems when compared to baselines of considering (i) only a mirror array-based \gls{RIS}, and (ii) a wall-LC-assisted VLC system. Also, the proposed deployment reveals that it is possible to improve the data rate and the energy efficiency of \gls{VLC} systems without using additional bandwidth resources and transmit power. 

Some interesting future research areas include, (i) extending the proposed system to a hybrid \gls{RF}/optical wireless communications (OWC), where \gls{RF} communications can be utilized in the uplink to support a complete uplink-downlink system, (ii) extending the proposed system to a multiple APs deployment with multi-users, while considering a static/dynamic \gls{RIS} configuration~\cite{9801736}, (iii) investigating the use of a laser diode (LD) instead of the white \gls{LED} used in the proposed system, (iv) validating the proposed theoretical system using an experimental test-bed, and (v) considering the illumination constraints and the human factors of the proposed \gls{VLC} system.
\vspace{-1.5em}
\bibliographystyle{IEEEtran}
\bibliography{main}


\begin{IEEEbiography}[{\includegraphics[width=1in,height=1.25in, clip,keepaspectratio]{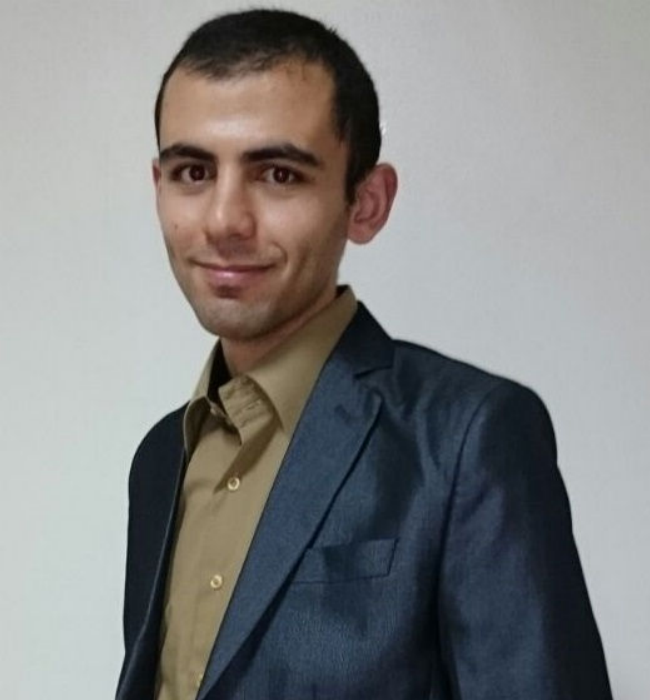}}]{\textbf{Omar Maraqa}} has received his B.S. degree in Electrical Engineering from Palestine Polytechnic University, Palestine, in 2011, his M.S. degree in Computer Engineering from King Fahd University of Petroleum \& Minerals (KFUPM), Dhahran, Saudi Arabia, in 2016, and his Ph.D. degree in Electrical Engineering at KFUPM, Dhahran, Saudi Arabia, in 2022. He is currently a Postdoctoral Research Fellow with the Department of Electrical and Computer Engineering, at McMaster University, Canada. His research interests include performance analysis and optimization of wireless communications systems.
\end{IEEEbiography}


\begin{IEEEbiography}[{\includegraphics[width=1in,height=1.25in, clip,keepaspectratio]{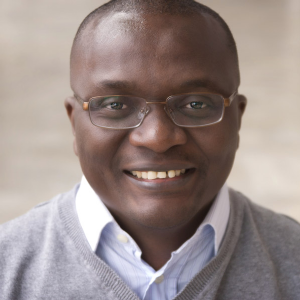}}]{\textbf{Telex M. N. Ngatched}} (M'05–SM'17) received the B.Sc. degree and the M.Sc. degree in electronics from the University of Yaoundé, Cameroon, in 1992 and 1993, respectively, the MscEng (Cum Laude) in electronic engineering from the University of Natal, Durban, South Africa, in 2002, and the Ph.D. in electronic engineering from the University of KwaZulu-Natal, Durban, South Africa, in 2006. From July 2006 to December 2007, he was with the University of KwaZulu-Natal as Postdoctoral Fellow, from 2008 to 2012 with the Department of Electrical and Computer Engineering, University of Manitoba, Canada, as a Research Associate, and from 2012 to 2022 with Memorial University. He joined McMaster University in January 2023, where he is currently an Associate Professor. Dr. Ngatched serves as an Area Editor for the IEEE Open Journal of the Communications Society, an Associate Technical Editor for the IEEE Communications Magazine, and an Editor of the IEEE Communications Society On-Line Content. He was the publication chair of the IEEE CWIT 2015, the Managing Editor of the IEEE Communications Letters and IEEE Communications Magazine, an Associate Editor with the IEEE Open Journal of the Communications Society and IEEE Communications Letters IEEE Communications Letters, the co-chair of the Spectrum Management, Radio Access technology, Services and Security track of VTC2021-Spring and VTC2022-Spring, and Technical Program Committee (TPC) member and session chair for many prominent IEEE conferences including GLOBECOM, ICC, WCNC, VTC, and PIMRC. He was a recipient of the Best Paper Award at the IEEE Wireless Communications and Networking Conference (WCNC) in 2019. He is a Professional Engineer (P. Eng.) registered with the Professional Engineers Ontario, Toronto, ON, Canada. His research interests include 5G and 6G enabling technologies, optical wireless communications, hybrid optical wireless and radio frequency communications, artificial intelligence and machine learning for communications, and underwater communications.
\end{IEEEbiography}

\end{document}